\documentclass[11pt]{article}
\usepackage[]{sao1}
\usepackage{graphicx}

\begin{document}

\title{Discovery of the shortest rotational period, non-degenerate, magnetic massive star by the MiMeS Collaboration}
\author{J.H. Grunhut\inst{1} \and  G.A. Wade\inst{1} \and T. Rivinius\inst{2} \and W.L.F. Marcolino\inst{3} \and R. Townsend\inst{4} \and and the MiMeS Collaboration}
\institute{Department of Physics, Royal Military College of Canada, Kingston, Ontario, Canada \and ESO - European Organisation for Astronomical Research in the Southern Hemisphere, Santiago, Chile \and LAM-UMR, CNRS \& Univ. de Provence, Marseille, France \and Department of Astronomy, University of Wisconsin-Madison, Madison, WI, USA}

\maketitle 

\begin{abstract}
We discuss the recent detection of a strong, organized magnetic field in the bright, broad-line B2V star, HR\,5907, using the ESPaDOnS spectropolarimeter on the CFHT as part of the Magnetism in Massive Stars (MiMeS) survey. We find a rotational period of 0.50833 days, making it the fastest-rotating, non-degenerate magnetic star ever detected. Like the previous rapid-rotation record holder HR 7355 (also discovered by MiMeS: Oksala et al. 2010, Rivinius et al. 2010), this star shows emission line variability that is diagnostic of a structured magnetosphere.
\end{abstract}

\section{Introduction} 
Magnetic fields are unexpected in hot, massive stars due to the lack of convection in their outer envelopes. However, a small number of massive B stars host strong, organized magnetic fields, such as the chemically peculiar He strong stars like the archetypical star $\sigma$~Ori~E and the recently discovered B2V star HR~7355 (Oksala et al. 2010; Rivinius et al. 2010). These stars are rapidly rotating and host strong magnetic fields that are coupled to a co-rotating magnetosphere (Townsend, Owocki, Groote 2005).

\section{Observations}
Twenty-one high-resolution ($R\sim68000$) observations of the variable B2V star HR\,5907 were obtained with the ESPaDOnS spectropolarimeter at the Canada-France-Hawaii telescope between February and August 2010. These initial observations clearly show the presence of Zeeman signatures in the circular polarization, Stokes $V$ Least-Squares Deconvolved (LSD), mean line profiles, indicative of a magnetic field.

We have also obtained six low-resolution ESO-VLT FORS polarimetric observations in addition to 27 UVES observations in April 2010 to follow up the ESPaDOnS observations and to further study the line-profile variability. 

\section{Rotational period and Fundamental Parameters}

\begin{figure}
\centering
\includegraphics[width=6.25in]{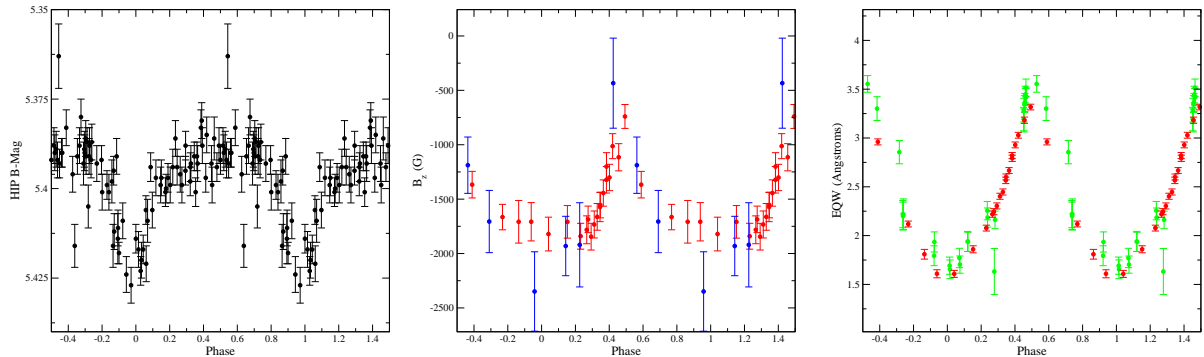}
\caption{Phased Hipparcos (left), longitudinal magnetic field (middle), and H$\alpha$ equivalent width measurements (right) for HR\,5907. Different colours correspond to the different instruments (red=ESPaDOnS, blue=FORS, green=UVES).}
\label{sidebyside}
\end{figure}

Using Hipparcos photometry, we find a single-wave period of $0.50831\pm0.00002$\,d (see Fig.~\ref{sidebyside}), which differs from the double-wave photometric light curve of HR\,7355 or $\sigma$~Ori E, likely indicating a different geometry of the magnetic field. From our longitudinal magnetic field measurements, we confirm this period, which we take to be the rotational period of this star (see Fig.~\ref{sidebyside}), making it the fastest rotating, non-degenerate, magnetic massive star! However, the period is sufficiently imprecise that the relative phasing between our current data and the Hipparcos data can be offset by so much as 0.5 cycles. Therefore, we have adopted a period of 0.50833\,d so that the relative phasing between the photometric minimum and the peak of the longitudinal field curve differs by 0.5 cycles - consistent with the predictions of semi-analytical models for a rotationally supported magnetosphere (Townsend 2008; Townsend \& Owocki 2005).

\begin{figure}
\centering
\includegraphics[width=5in]{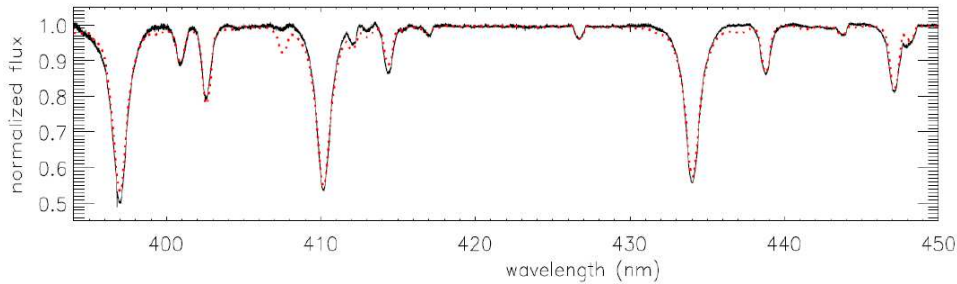}
\caption{Comparison between our ESPaDOnS spectrum (black) with our best-fit NLTE model fit (dotted-red).}
\label{nlte_fit}
\end{figure}

From NLTE model fits to the ESPaDOnS spectra (shown in Fig.~\ref{nlte_fit}), we find that $T_{\rm eff}=19\pm2$\,kK, $\log(g)=2.95\pm0.04$, and $v\sin i=280\pm10$\,km\,s$^{-1}$.

\section{Line Profile Variability}
\begin{figure}
\centering
\includegraphics[width=2.8in]{grunhut_hr5907_fig3.eps}
\includegraphics[width=2.8in]{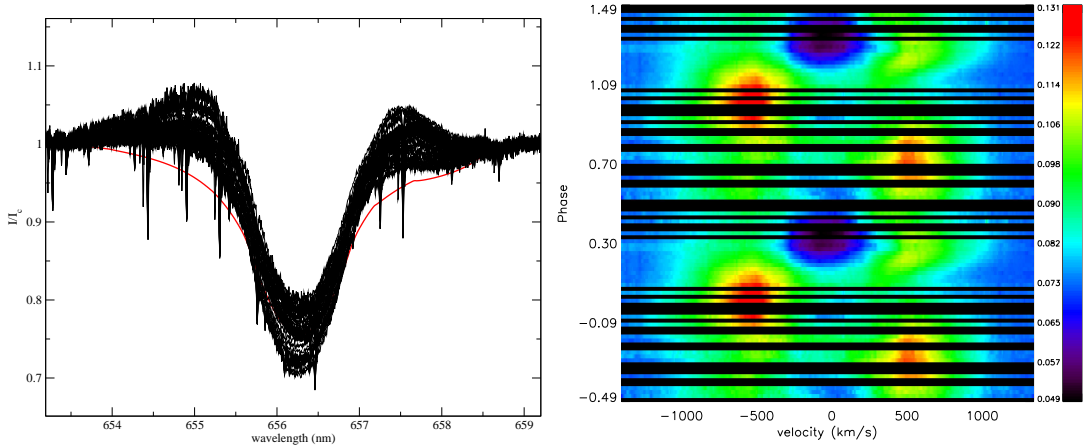}\\
\caption{{\bf Left:} H$\alpha$ (black) profiles for different nights compared to a NLTE model profile (red). {\bf Right:} Phased residual H$\alpha$ variations relative to the NLTE model.}
\label{ha_dyn}
\end{figure}

In addition to the photometric and magnetic periodicity, we also find that H$\alpha$ varies with the same period, as illustrated by the equivalent width variations shown in Fig.~\ref{sidebyside}. H$\alpha$ shows line profile variations of emission extending to high velocities, as shown in Fig.~\ref{ha_dyn}. The double-lobed pattern and equivalent width variations strongly suggests that HR\,5907 hosts a structured magnetosphere similar to $\sigma$~Ori~E and HR\,7355 consisting of co-rotating, magnetically confined clouds of stellar wind plasma.

\section{Magnetic Field Geometry}
\begin{figure}
\centering
\includegraphics[width=5.2in]{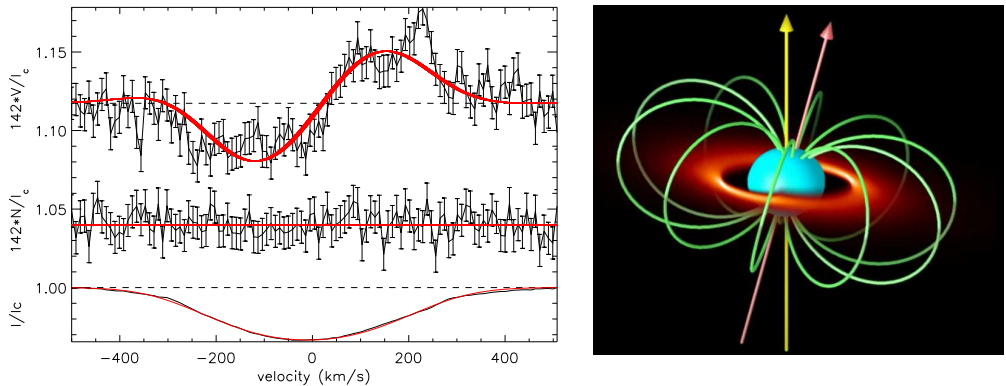}
\caption{{\bf Left:} Example dipole model fit, characterized by  $B_d=18$\,kG and $\beta=4^{\circ}$ (red) to the observed LSD Stokes $V$ (top), diagnostic null (middle) and unpolarized Stokes $I$ profile (bottom) of HR\,5907 (black). {\bf Right:} Illustration of the magnetic field and circumstellar environment for a star with a rotation rate and magnetic field geometry similar to HR\,5907 (Image courtesy of R.H.D. Townsend).}
\label{lsd_fits}
\end{figure}

Assuming rigid rotation, we infer that the inclination $i\sim75^{\circ}$. From fits to the longitudinal field curve shown in Fig.~\ref{sidebyside} we estimate that HR\,5907 hosts a mainly dipole magnetic field, with a strength at its pole of $\sim20$\,kG, and a magnetic axis nearly aligned with the rotation axis. Preliminary fits of a dipole model to the LSD profiles at each individual phase seems to confirm this estimate, as shown in Fig.~\ref{lsd_fits}. Also included in Fig.~\ref{lsd_fits} is an illustration of how the magnetic field and circumstellar environment of a star with a similar rotation rate and similar magnetic field configuration would appear edge-on. Currently, the inclination is poorly constrained, which results in a large possible range for the dipole field strength. However, using the predictions of the Rigidly Rotating Magnetosphere model (Townsend \& Owocki 2005), we expect to better constrain the geometry of HR\,5907 based on the variations shown in Fig.~\ref{sidebyside}.

\end{document}